\documentclass{article}

\usepackage{arxiv}

\usepackage[utf8]{inputenc} 
\usepackage[T1]{fontenc}    
\usepackage{hyperref}       
\usepackage{url}            
\usepackage{booktabs}       
\usepackage{amsfonts}       
\usepackage{nicefrac}       
\usepackage{microtype}      
\usepackage{lipsum}		
\usepackage{graphicx}
\usepackage{natbib}
\usepackage{doi}
\usepackage{subcaption}
\usepackage{multicol, multirow}

\title{MAPLES-DR: MESSIDOR Anatomical and Pathological Labels for Explainable Screening of Diabetic Retinopathy}

\date{December 15, 2023}	

\author{ \href{https://orcid.org/0009-0000-4110-8227}{\includegraphics[scale=0.06]{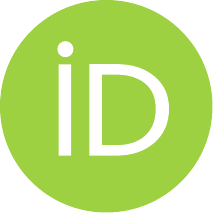}\hspace{1mm}Gabriel Lepetit-Aimon} \\
	Department of Informatics\\
	Polytechnique Montréal\\
	Montréal, QC H3T 1J4, Canada \\
	\texttt{gabriel.lepetit-aimon@polymtl.ca} \\
	\And
	\href{https://orcid.org/0000-0002-0978-3588}{\includegraphics[scale=0.06]{orcid.pdf}\hspace{1mm}Clement Playout} \\
	Department of Informatics\\
	Polytechnique Montréal\\
	Montréal, QC H3T 1J4, Canada \\
	\texttt{clement.playout@polymtl.ca} \\
        \And
        Marie Carole Boucher \\
        Faculté de Médecine, Département d'ophtalmologie \\
        Université de Montréal \\
        Montréal, QC H3T 1J4, Canada \\
	\texttt{mariecarole@gmail.com} \\
        \And
        Renaud Duval \\
        Faculté de Médecine, Département d'ophtalmologie \\
        Université de Montréal \\
        Montréal, QC H3T 1J4, Canada \\
	\texttt{renaud.duval@gmail.com} \\
        \And
        Micheal H. Brent \\
        University Health Network,
        Toronto, Canada \\
        \texttt{mhbrentmd@gmail.com}\\
        \And
	\href{https://orcid.org/0000-0001-6170-5627}{\includegraphics[scale=0.06]{orcid.pdf}\hspace{1mm}Farida Cheriet} \\
	Department of Informatics\\
	Polytechnique Montréal\\
	Montréal, QC H3T 1J4, Canada \\
	\texttt{clement.playout@polymtl.ca} \\
}

\renewcommand{\shorttitle}{MAPLES-DR}

\hypersetup{
pdftitle={MAPLES-DR},
pdfsubject={q-bio.NC, q-bio.QM},
pdfauthor={Gabriel Lepetit-Aimon, Clément Playout, Marie Carole Boucher, Renaud Duval, Michael H Brent, Farida Cheriet},
}

\begin{document}
\maketitle

\begin{abstract}

Reliable automatic diagnosis of Diabetic Retinopathy (DR) and Macular Edema (ME) is an invaluable asset in improving the rate of monitored patients among at-risk populations and in enabling earlier treatments before the pathology progresses and threatens vision. However, the explainability of screening models is still an open question, and specifically designed datasets are required to support the research.

We present MAPLES-DR (MESSIDOR Anatomical and Pathological Labels for Explainable Screening of Diabetic Retinopathy), which contains, for 198 images of the MESSIDOR public fundus dataset, new diagnoses for DR and ME as well as new pixel-wise segmentation maps for 10 anatomical and pathological biomarkers related to DR.
This paper documents the design choices and the annotation procedure that produced MAPLES-DR, discusses the interobserver variability and the overall quality of the annotations, and provides guidelines on using the dataset in a machine learning context.

\end{abstract}


\clearpage

\section*{Background \& Summary}
Diabetic retinopathy (DR) is a complication of diabetes mellitus (DM) that damages the retinal microvasculature and can lead to vision impairment. DR develops gradually and is clinically characterized by stages according to the presence of lesions in the retina. Non-invasive fundus imaging can detect these lesions, and regular screening of at-risk populations is necessary to ensure early treatment and preserve their vision\cite{Lanzetta2020}. Despite widespread screening programs in North America\cite{Committee2013,Hooper2017,Egunsola2021}, 40\% of patients with DM are still not monitored for DR. Studies based on these programs indicate that teleophthalmology screening increases the rate of monitored patients \cite{Avidor2020}, and that algorithms for automatic diagnosis of DR can enable further improvements by reducing costs and increasing the frequency of examinations, as well as ensuring timelier management of referred patients \cite{Lanzetta2020}.

In the last decade, machine learning models have been successfully applied to the automatic diagnosis of DR in fundus images. 
At the core of the development of these supervised algorithms are public datasets of annotated images, used for training and validation\cite{raman2019fundus}. Among them, the Eyepacs\cite{diabetic-retinopathy-detection} and MESSIDOR-2\cite{MESSIDOR} datasets have enabled the development of automatic screening algorithms that exceed the performance required by the FDA \cite{raman2019fundus}, and even outperform human experts \cite{Gulshan2019}. Yet, medical personnel still lack confidence in these technologies, which do not meet the standards of explainable AI.
We believe that to overcome this limitation, improving training labels should not be overlooked. Indeed, uncertainty around validation methods and generalization issues call for more diverse and bias-aware validation datasets and better documentation of their labeling process. The weak explainability of screening models calls, in addition, for more exhaustive and clinically relevant training labels beyond simple diagnostic classification.
In its report ``Four principles of Explainable AI'', the National Institute of Standards and Technology underlines that a proper explanation of algorithm outcomes should be \emph{meaningful} to the target audience \cite{phillips2020four}, namely, in our case, it must be formulated using a vocabulary familiar to ophthalmologists. Restricting training labels to only DR grades is a major shrinkage of medical vocabulary, especially when clinical justifications rely heavily on identifying the anatomical and pathological structures of the retina (vessels, macula, red or bright lesions, etc.)\cite{wilkinson2003proposed, Zachariah2015, Boucher2020}.
Therefore, whether it is to guide models \textit{a priori} so that they learn representations compatible with clinical knowledge or to interpret these representations \textit{a posteriori} by comparing them with known biomarkers, datasets with pixel-wise annotations of such structures play a key role in the development of explainable screening models for DR.
We designed \emph{MAPLES-DR} (MESSIDOR Anatomical and Pathological Labels for Explainable Screening of Diabetic Retinopathy) with this objective in mind.

Annotating pixel-wise labels requires considerably more time and effort than image-level diagnostic labels; consequently, fundus datasets with such labels are few. Most of them publish annotations for only one biomarker (e.g. vessels\cite{DRIVE, HRF, FIVES, RETA}, optic disc\cite{REFUGE,PAPILA,CHAKSU}, exudates\cite{SUSTECH}, microaneurysms \cite{Decenciere2013}). To our knowledge, only four datasets provide labels for four lesions that are symptomatic of DR: microaneurysms, hemorrhages, exudates, and cotton-wool spots (CWS). FGADR \cite{FGADR} and Retinal Lesions \cite{RetinalLesions} comprise numerous images (1842 and 1593 respectively), but pathological structures are mostly annotated as large clusters of lesions. In contrast, IDRiD \cite{IDRID} and DDR \cite{DDR}  provides precise segmentations of each lesion, but for fewer images (81 and 757 images, respectively). The articles accompanying these datasets focus on the performance gain they bring to existing screening methods, but provide only shallow descriptions of the protocols used to produce their annotations. None of them contain annotations of anatomical structures.

\begin{figure}[ht]
    \centering
        \includegraphics[width=\textwidth]{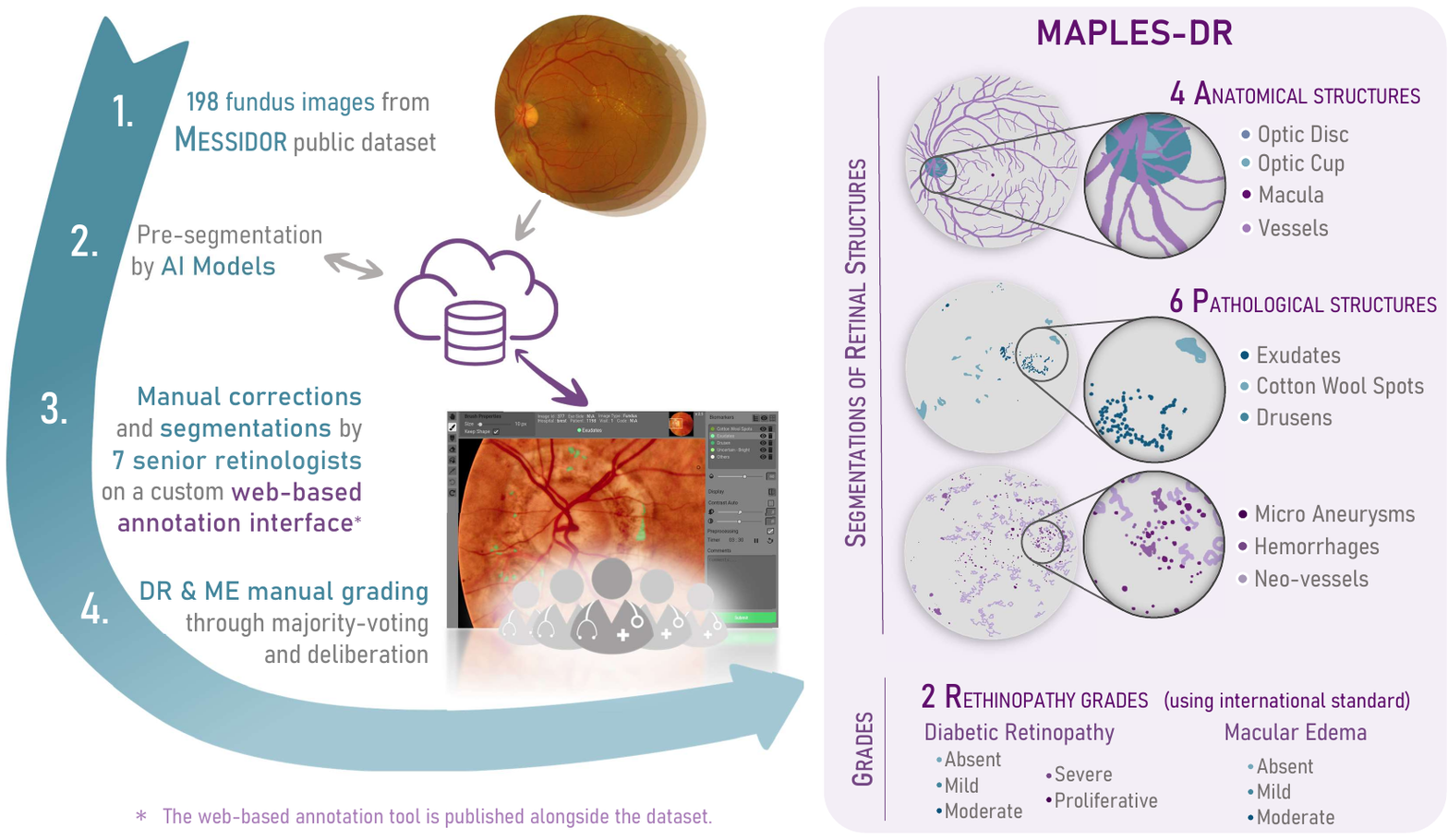}
    
  \caption{Schematic overview of the MAPLES-DR annotation protocol and of the content of the dataset.}
    \label{fig:GraphicalAbstract}
\end{figure}

The \emph{MAPLES-DR} dataset consists of new annotations for 198 fundus images from the MESSIDOR-2\cite{MESSIDOR} dataset. For each of these images, our team of senior ophthalmologists graded DR and macular edema (ME) and segmented 10 anatomical and pathological structures (see Figure~\ref{fig:GraphicalAbstract}).  As far as we know, no other public dataset offers such a comprehensive range of anatomical and pathological labels for fundus images. This paper extensively documents the design choices and labeling protocol of this dataset, as well as its potential biases and inaccuracies.
By thoroughly documenting our approach and releasing all the tools and software that produced the annotations along with this paper, we aim to help other research teams build more such datasets for explainable DR screening.

\section*{Methods}

\subsection*{Fundus images description}
MAPLES-DR does not provide new fundus images, but only \emph{new annotations} for those already published in 2014 by the MESSIDOR consortium. MESSIDOR fundus images are high resolution (between 1440~$\times$~960 and 2304~$\times$~1536 pixels) and were acquired at three French ophthalmology centers.
These images are not included in our repository, but are freely available to researchers who request them on the \href{https://www.adcis.net/fr/logiciels-tiers/messidor-fr/}{MESSIDOR Consortium's website}. It is under those terms that we have been able to access them and create annotations of our own.

In addition to fundus images, the MESSIDOR database also provides global DR and ME grades labeled by the consortium's clinicians. Diabetic retinopathy is graded from R0 (healthy) to R3 (proliferative) according to the number of microaneurysms, hemorrhages, and neovascularization, while macular edema is graded from M0 (healthy) to M2 according to the position of the exudate closest to the fovea. 

\begin{figure}[ht]
\centering
\includegraphics[width=0.7\linewidth]{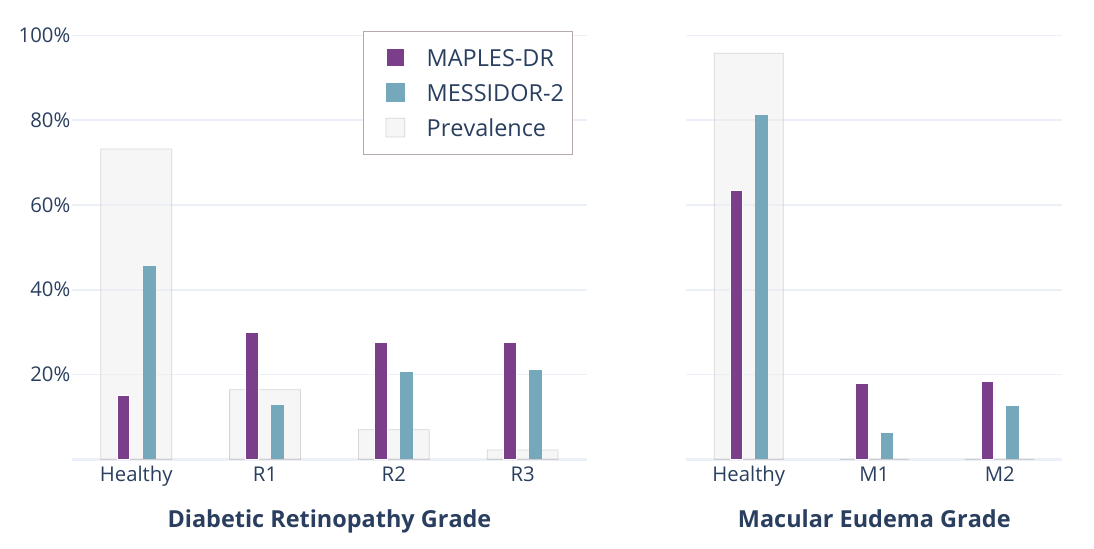}
\caption{Disparities among DR and ME grade distributions (converted to MESSIDOR grades) between MAPLES-DR subset (purple), MESSIDOR-2 complete dataset (teal blue) and prevalence from first screening of the Toronto tele-ophthalmology program\cite{Cao2022} (gray).}
\label{fig:DR-dist}
\end{figure}

Of the 1,200 images available in the MESSIDOR dataset, 200 were selected for annotation in MAPLES-DR. We randomly chose 30 healthy images, 59 R1 images, 55 R2 images, and 56 R3 images, according to the grades provided by MESSIDOR. This selection is not representative of the prevalence of the pathology in either the MESSIDOR database or an actual screened population (see Figure~\ref{fig:DR-dist}). However, it provides a set of images for each pathology stage and focuses on those critical for screening (R1 and R2). In the latest stage of the annotation campaign, we found two duplicate entries in the dataset (MESSIDOR-2 contains several duplicate images stored with different names). We excluded them from MAPLES-DR, reducing the number of images to 198. Still, we publish their annotations in a separate folder as an example of intraobserver variability in labeling.

\subsection*{DR and ME grade description}
The diagnoses provided with the MESSIDOR database for DR and ME are based on an uncommon grading system. To facilitate comparison with other public fundus image databases, retinologists regraded the images following the guidelines for Canadian teleopthalmology screening, whose grades are closer to international standards (i.e. ICDR\cite{wilkinson2003proposed} or the Scottish system \cite{Zachariah2015}). 

These guidelines distinguish six grades for DR (R0: absent, R1: mild, R2: moderate, R3: severe, R4A: proliferative, R4S: stable treated proliferative) and three for ME (M0: absent, M1: mild, M2: moderate). Each grade is associated with a recommended course of action (from rescreening in 12-24 months to immediate referral to an ophthalmologist) and is determined by the number and position of red lesions (hemorrhages, microaneurysms, neovessels) and bright lesions (exudates, cotton wool spots) visible in the retina. A comprehensive definition is available in the literature\cite{Boucher2020}.

\subsection*{Retinal biomarker description}
The main contribution of MAPLES-DR is the annotation of retinal structures symptomatic of DR at a pixel-level. The decision to include or exclude retinal structures from the list of annotated biomarkers was based on clinical understanding of their role in DR histopathology and screening procedures.

\subsubsection*{Anatomical structures}
Anatomical structures are present in all images, including healthy ones, but their appearance and their proximity to lesions provide valuable diagnostic information.

\emph{Retinal vessels} are indicative of the stage of DR: an increase in arteriolar tortuosity is associated with mild and moderate stages \cite{sasongkoRetinalVascularTortuosity2011}, while venous beading and dilation are symptoms of severe proliferative stages. The vascular tree is also used as a reference to assess the readability of an image.

The \emph{optic disc}, \emph{optic cup}, and \emph{macula} are also included in MAPLES-DR. Their purpose for diagnosis is two-fold. \textbf{1.}~ME is graded by counting the number of lesions within one or two optic disk diameters from the macula, which implies the annotation of both these anatomical structures. Similarly, clinical definitions of DR severity often distinguish four quadrants by diving the retina  horizontally by a line through the fovea and optic disc (superior / inferior division) and vertically by a line through the fovea (temporal / nasal division) \cite{purvesRetinotopicRepresentationVisual2001}. \textbf{2.}~The positions of the lesions in relation to these healthy structures may indicate different etiologies and severities. For example, clinical guidelines sometimes distinguish between disc neovascularization and other neovascularization.

\subsubsection*{Red lesions}
Diabete mellitelus affects the walls of the vessels, eventually causing microvascular dysfunctions that manifest in the retina as microaneurysms, hemorrhages, intraretinal microvascular abnormalities (IRMA), or neovessels. We refer to these pathological structures as "red lesions". 

\emph{Microaneurysms} appear as small circular dilations of the capillaries. They are early signs of microvascular dysfunction and are commonly used to detect mild DR.

\emph{Intraretinal hemorrhages} develop in more advanced stages of the pathology and are divided into dot or blot hemorrhages. Dot hemorrhages appear as circular and well-defined spots and are typically caused by the rupture of a microaneurysm. Distinguishing them from microaneurysms is challenging, and only fundus angiography (FA) can differentiate the two with complete certainty. Blot hemorrhages are larger and have less defined borders. Both were annotated simply as \emph{hemorrhages} in MAPLES-DR. Clinical practice also recognizes superficial (flame-shaped) and vitreous hemorrhages that appear in the most severe stages of retinopathy. But because none was discovered in the MAPLES-DR dataset, they were excluded from the biomarker list.

Starting from the moderate non-proliferative stage (R2), irregular intraretinal vessels can appear, referred to as \emph{IRMA}. The next stage of the disease (R3) coincides with even more extensive intraretinal changes, which are precursors to worsening of the disease. Indeed, the presence of IRMA indicates a 50\% risk of developing \emph{neovascularisation} (NV) within one year, corresponding to a transition to the proliferative stage of the disease. Leakages from extensive NV are responsible for preretinal and vitreous hemorrhages that can cause major visual loss. In the fundus image, NVs are difficult to distinguish from IRMA; however, fluorescein angiography may reveal a leakage that serves as a discriminant factor between the two. In the absence of this imaging modality, IRMA are not differentiated from NV in MAPLES-DR.

\subsubsection*{Bright Lesions}
In the severe stages of DR, the retina thickens (edema formation) and hard \emph{exudates} (also known as lipoprotein exudation)  may appear, potentially causing loss of visual acuity. These deposits usually arise from leakage from damaged capillaries.
Furthermore, in the case of ischemia, one can observe a blockage in axonal transport (the movement of mitochondria, lipids, proteins, and other substances within the neuron's body, allowing for its renewal) in the optic nerve fiber layer. This can lead to the appearance of lesions known as \emph{Cotton Wool Spots} (CWS), resulting from axoplasmic accumulations. They are characterized by their white appearance and blurry borders. While the principal etiology is diabetic retinopathy, CWS can also be observed in other vascular diseases (systemic artial hypertension, vein obstruction, coagulopathies...) 
\\
Finally, we provide annotations of \emph{drusens}. These lesions are more commonly associated with Age-related Macular Degeneration (AMD), with a prevalence varying from 10\% (fifth decade of life) to 35\% (seventh decade).  They usually appear around the macula and are histologically situated at the interface with the Retinal Pigment Epithelium (RPE). It is supposed that they originate from degenerative products of the RPE's cells and are composed of lipids and glycoproteins. Classifying early stage AMD depends on  estimating the size of the drusen.

\subsection*{Preparation and timeline of the annotation campaign}
MAPLES-DR was annotated by seven Canadian retinologists affiliated with five different hospitals: Hôpital Maisonneuve Rosemont, Montreal; University Health Network, Toronto; Centre Hospitalier Universitaire de Montréal; Université de Sherbrooke; and Centre Hospitalier Universitaire Saint-Justine, Montreal). All annotators were seniors retinologists and were recruited thanks to their involvement in teleophthalmology programs for the detection of DR in Quebec and Ontario. 

The annotation procedure was co-designed with these clinicians to meet a triple objective: \textbf{1.}Providing an intuitive yet effective annotation tool for the classification and segmentation of biomarkers in fundus images. \textbf{2.} Enabling a collaborative effort on common annotations despite the geographical distance between the retinologists and the limited time each could dedicate to this program. \textbf{3.} Designing a "scalable" annotation protocol, capable of being extended to much more ambitious annotation campaigns, such as labeling large Canadian telemedicine databases containing tens of thousands of images.

To meet these challenges, we developed a custom web-based annotation platform allowing the following workflow: expert annotators can access the Web portal at any time to consult and edit annotations with specialized drawing tools; these annotations and the related information (annotation times, comments) are centralized and stored in a secure database hosted on our laboratory server; as the research team, we assign tasks to annotators, monitor progress, and export annotations via a Python API. The annotation platform (portal, annotation tools, server backend, and Python API) as well as training material for annotators are published with this article (see Code Availability section).

The platform was designed and implemented during fall 2018 and tested by the annotation team in January 2019. This phase was also an opportunity to train annotators in using the annotation tools through written documentation and, for some, videoconference workshops. The biomarker annotation campaign began in February 2019 and ended in February 2020. Throughout this period, dialogue with the annotation team led to improvements of the annotation tools (refining presets, adjusting visualization enhancement tools, correcting the behavior of drawing tools, etc.) and to harmonizing the annotation practices.
The annotation campaign for grading of DR and ME was carried out in a second phase, in September and October 2020.

\subsection*{Procedure to annotate retinal structures }
Labeling anatomical and pathological retinal structures at a pixel level is an intrinsically tedious task. To limit the time and commitment required of retinologists, we implemented an efficient procedure to label retinal structures. Retinologists were not asked to annotate these structures from scratch but instead were provided with segmentation maps preannotated by AI models.

\subsubsection*{Reviewing and correcting pre-annotated segmentation maps}
The MAPLES-DR annotation project was initially motivated by the scarcity of publicly available datasets. This restricted the performance of AI models designed to segment retinal structures, and precluded their clinical use. Fortunately, these models were still capable of producing satisfactory initial segmentations to facilitate manual annotation. Hence, instead of labeling from scratch, retinologists were assigned the less laborious task of reviewing AI-generated maps (we will refer to them as \emph{preannotated maps}) and correcting any segmentation errors. This included removing false positive lesions, adding lesions that the algorithm had missed (false negatives), and adjusting the boundaries of some lesions or changing their type (e.g. from microaneurysm to hemorrhage).
This preannotated map review process was used to segment exudates, microaneurysms, hemorrhages, and blood vessels. Labeling the other anatomical biomarkers (macula, optic disc, and optic cup) was more straightforward and didn't require preannotation. As for the other pathological biomarkers (neovessels, drusen and CWS), the databases published at the time were not sufficient to train segmentation models and therefore those biomarkers were labeled from scratch.

In practice, generating the preannoted maps was entrusted to two models: one responsible for segmenting lesions, the other for segmenting vessels. Both used UNet-based architectures refined to better meet the challenges of their respective tasks. The lesions model used weak supervision and multitasking\cite{playout2019novel} to predict two segmentation maps for bright and red lesions. The predicted bright lesions were used as exudate preannotations and the red lesions segmentation map was divided into microaneurysms and hemorrhages according to their size. The vessel model included a low-resolution branch to efficiently increase its receptive field and a conditional random fields to improve its segmentation of small vessels\cite{Lepetit2018}. Both models were state of the art in 2018. The final segmentation thresholds of both models were slightly reduced to favor false positives over false negatives. Indeed, we anticipated that reviewing all preannotated structures and removing the incorrect ones would be an easier task than spotting and annotating all those missing. Conversely, no lesions were pre-annotated for healthy images (according to MESSIDOR's experts), to avoid the need of removing them.

Upon reaching 100 reviewed and corrected vessel segmentation, the weights of the pre-annotation model were fine-tuned on those new labels. The resulting improvements to the automated segmentation maps further reduced by 10\% the average time required to annotate vessels.

\subsubsection*{Labeling instructions provided to retinologists}
In preparation of the annotation process, we separated the 10 biomarkers into 4 categories: \textbf{1.} red lesions (micro-aneurysms, hemorrhages, and neo-vessels), \textbf{2.} bright lesions (exudates, drusen, cotton wool spots), \textbf{3.} vessels, and \textbf{4.} other anatomical biomarkers outside vessels (optic disc, optic cup, fovea). An additional "uncertain" channel was added to the two pathological biomarker categories (i.e. \emph{uncertain bright} and \emph{uncertain red}) in anticipation of ambiguous cases. Here, annotators could annotate any structures that appeared pathological, but whose exact nature was unclear. 

Each retinologist was assigned one of four categories and focused on annotating only the corresponding biomarkers. Consequently, each of the 198 images was labeled by several retinologists: one segmenting vessels, another the red lesions, etc. By splitting the annotation tasks as such and specializing each clinician in the labeling of a few biomarkers, we intended to simplify the learning curve for the annotation tools and speed up the entire annotation process. This measure, combined with preannotated labeling, reduced the cumulative time needed to annotate an image to an average of 22 minutes.

Besides the list of biomarkers to label and a list of recommendations on using the annotation tools, no explicit labeling instructions were provided to clinicians. We trusted their diagnostic expertise and did not provide instructions on the clinical definition of biomarkers or the level of detail expected. Yet, it is worth noting that the platform's design induces implicit guidelines. Lesions in the preannotated maps were small (the median lesion diameter was 13 pixels) and numerous; this influenced the annotators to label each lesion individually (instead of circling the general area) and to pay attention to all lesions, even the smallest ones. As Radsch et al. \cite{radsch2023labelling} recently reported, providing expert annotators with exemplary images instead of extended text descriptions significantly improves the quality of biomedical annotations.

\subsubsection*{Particularities of the segmentation annotation tools}
When designing the annotation tools, our focus was on ergonomics and annotation efficiency. However, certain design choices had more impact on the final annotations than anticipated. We report them here to point out possible biases in the MAPLES-DR annotations and to encourage other teams implementing such tools to pay attention to them.

Two image pre-processing algorithms  were included as visualization enhancement tools to assist annotators in their diagnosis. The first one corrected illumination variations while preserving image hues, by subtracting a median filter and using contrast-limited adaptive histogram equalization (CLAHE) on the brightness channel only. The second one maximized the contrast by independently normalizing the intensities of each RGB channel so that they spread over their full range of values. These two processes could be activated individually or in combination (see Figure~\ref{fig:visualisationTools}).
The decision of how and when to use these image enhancements was left to the clinician's discretion. Their feedback indicated that CLAHE pre-processing was particularly appreciated for spotting small red lesions (i.e. microaneurysms and small hemorrhages) at first glance.  However, it should be noted that while this treatment preserves general hues, it still modifies the natural color signature of retinal structures. This can affect the user's ability to distinguish microaneurysms from hemorrhages or even image artifacts (e.g. dust specks). As a result, the labels published in MAPLES-DR might have a higher detection rate for red lesions than other databases annotated without such pre-processing.

\def\colSize{.245\textwidth}
\begin{figure}[ht]
    \centering
    \begin{subfigure}{\colSize}
        \includegraphics[width=\textwidth]{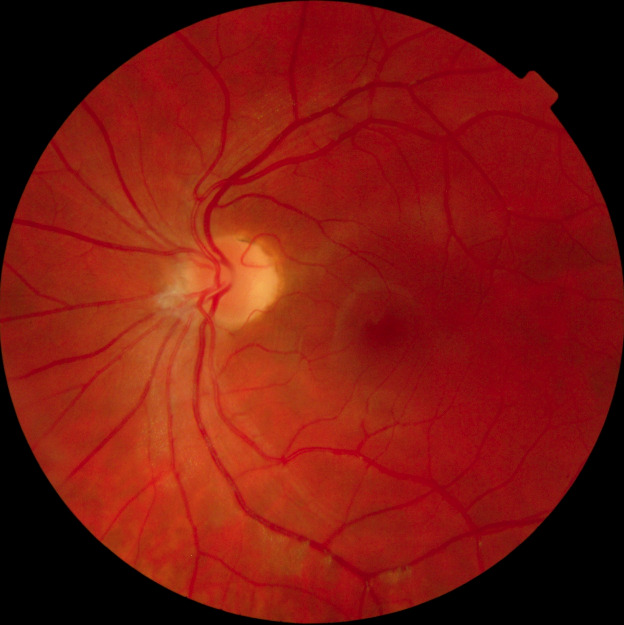}
        \subcaption{No pre-processing}
    \end{subfigure}
    \begin{subfigure}{\colSize}
        \includegraphics[width=\textwidth]{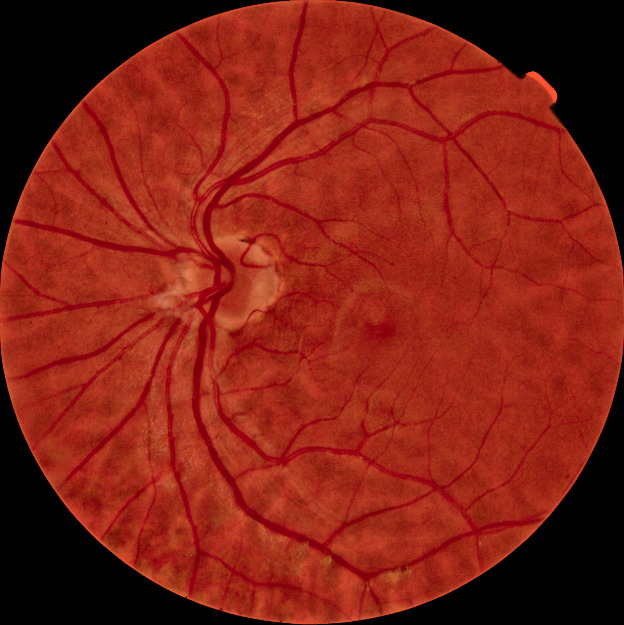}
        \subcaption{CLAHE}
    \end{subfigure}
    \begin{subfigure}{\colSize}
        \includegraphics[width=\textwidth]{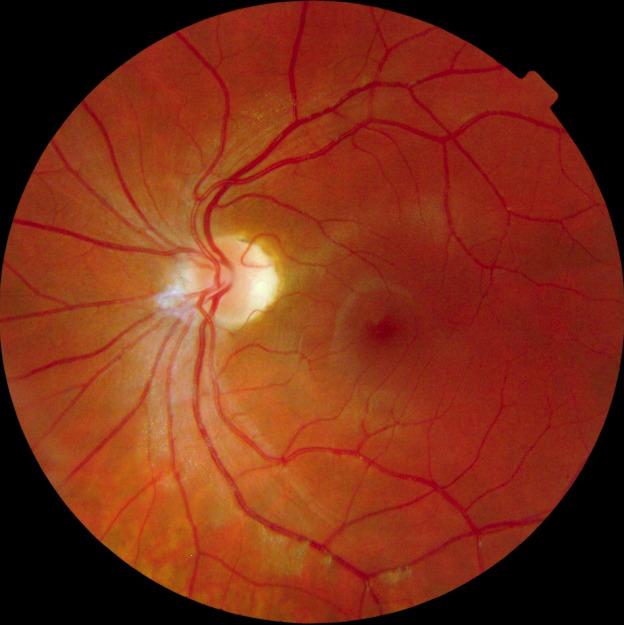}
        \subcaption{Contrast Maximization}
    \end{subfigure}
    \begin{subfigure}{\colSize}
        \includegraphics[width=\textwidth]{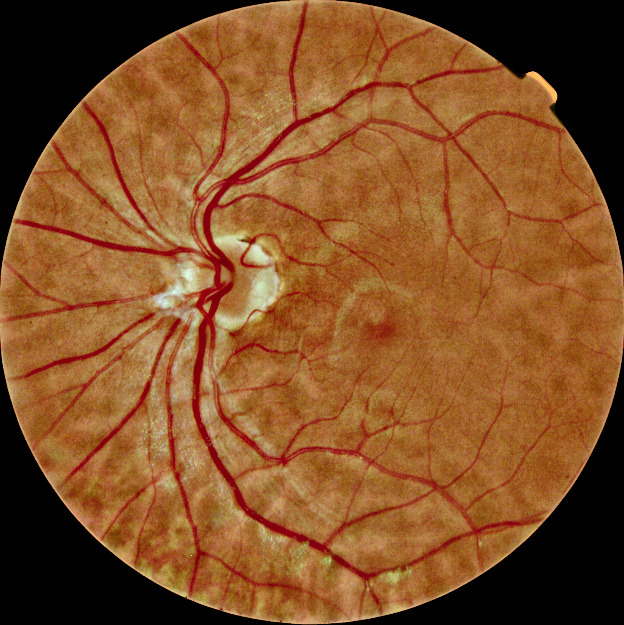}
        \subcaption{CLAHE + Contrast}
    \end{subfigure}
    
  \caption{Fundus image pre-processed with the visual enhancement algorithms available in the annotation platform.}
    \label{fig:visualisationTools}
\end{figure}

The default configuration of the segmentation tools was also identified as a source of potential annotation bias. When annotating small lesions (e.g. microaneurysms), the annotators tended to keep the default brush size of 10 px, resulting in an abnormally high number of lesions with precisely this size. To mitigate this bias, we implemented a pressure-sensitive brush for clinicians using a graphics tablet that dynamically adjusts the brush size based on the pressure applied to the stylus. This feature has proven to be especially useful for vessel annotation.

\subsubsection*{Post-processing}
The retinologists' manual annotations underwent a few post-processing steps to reach their final form published in MAPLES-DR. 

Before annotation, all fundus images were cropped and rescaled to a resolution of 1500$\times$1500 pixels. This was necessary to ensure a consistent level of segmentation detail across all images. But it also hinders the use of the dataset in a machine learning context because it introduces a resolution mismatch between MAPLES-DR segmentation maps and their corresponding fundus images in the MESSIDOR dataset. Therefore, all annotated maps were spatially padded and resampled using nearest-neighbor interpolation to match the sizes of the original images. The segmentation maps at their initial resolution of 1500$\times$1500 pixels are included in the \texttt{biomarkers-supplementaryInformation.zip} archive.

\paragraph{Stratification process}
To favor reproducible results for algorithms using MAPLES-DR, we split the dataset into training and  test sets. These contain, respectively, 138 and 60 images (70\% / 30\%). To divide the dataset, we built a multilabel vector for each image indicating the presence or absence of each lesion type. The sets were then obtained by applying iterative multilabel-shuffle stratification as introduced by Sechidis et al. \cite{sechidisStratificationMultilabelData2011}.

\subsection*{Procedure to grade DR and ME }
Three retinologists were asked to grade the images according to Canadian Teleophtalmic Guidelines\cite{Boucher2020} (R0-R4 and M0-M2) and could also flag images of insufficient quality for diagnosis. 

The variability between observers in the diagnosis of DR and ME is not negligible and is well documented in the literature \cite{Ruamviboonsuk2006, bragge2011, shi2015, Krause2018,teoh2023variability}. 
Our DR and ME grading procedures accounted for this by splitting diagnosis labeling into two phases. Each image was individually graded by three senior retinologists and a common diagnosis was determined by majority voting. The few cases of complete disagreement between the three retinologists (5 cases of DR and 12 cases of ME out of 198 images) were resolved in a second phase by deliberation between the experts to establish a consensual diagnosis. In the interest of transparency, we provide both the individual diagnoses collected during the first phase and the final diagnoses resulting from the deliberation of the second phase (when required).

\section*{Data Records}
The MAPLES-DR dataset is available at \href{https://doi.org/10.6084/m9.figshare.24328660}{https://doi.org/10.6084/m9.figshare.24328660} and contains four files: \texttt{biomarkers-\allowbreak Multi\allowbreak Classes\allowbreak Maps.zip}, \texttt{biomarkers-\allowbreak Multi\allowbreak Labels\allowbreak Maps\allowbreak.zip}, \texttt{diagnosis.xls} and \texttt{biomarkers-\allowbreak supplementary\allowbreak Information.zip}. 

\paragraph{biomarkers-MultiLabelsMaps.zip} contains a \texttt{train/} and a \texttt{test/} folder. Each consists of 12 subfolders (one for each biomarker), which in turn hold the segmentation maps resulting from the labeling of pathological and anatomical biomarkers by our team of retinologists. 138 training maps and 60 testing maps are provided as such for each biomarker. Segmentation maps are encoded in binary PNG format, where pixels belonging to the biomaker are set to \texttt{1} while those belonging to the background are set to \texttt{0}. The images were rescaled and renamed to match their initial resolution and naming convention in the MESSIDOR dataset.

\paragraph{biomarkers-MultiClassesMaps.zip} provides the same data as \texttt{segmentationMaps.zip}, but in a format more appropriate for training semantic segmentation models. The \texttt{train/} and \texttt{test/} folders consists of only five subfolders corresponding to the biomarker categories: \texttt{All/}, \texttt{BrightLesions/}, \texttt{RedLesions/}, \texttt{Vessels/} and \texttt{OtherAnatomical/}. Each subfolder contains multilabel maps whose pixel values follow the label definition presented in Table~\ref{tab:semanticSegmentationLabels}. For pixels belonging to multiple biomarkers, the label of higher value always has precedence. The maps are saved as 8-bit single-channel PNG images with the same resolution and naming convention as the MESSIDOR dataset.

\begin{table}
    \centering
    \begin{tabular}{@{}rllll@{}}
    \toprule
    &\multicolumn{4}{c}{\textbf{Multiclasses Semantic Segmentation Maps}}\\ 
                               Label &\texttt{OtherAnatomical}&  \texttt{Vessels}&\texttt{BrightLesions}& \texttt{RedLesions}\\
                                \midrule
    \texttt{0} &Background (\texttt{0})&Background (\texttt{0})&Background (\texttt{0})&Background (\texttt{0})\\
 \texttt{1}& Optic Disc (\texttt{1})& Vessels (\texttt{4})& Uncertain Bright (\texttt{5})&Uncertain Red (\texttt{9})\\
    \texttt{2}&Optic Cup (\texttt{2})&  &Exudates (\texttt{6})& Neovascularization (\texttt{10})\\
    \texttt{3}&Macula (\texttt{3})&  &Cotton Wool Spot (\texttt{7})& Microaneurysms (\texttt{11})\\
    \texttt{4}&&  &Drusen (\texttt{8})& Hemorrhages (\texttt{12})\\
    \bottomrule
    \end{tabular}

    \caption{Definition of labels for semantic segmentation maps. The numbers in parentheses define the biomarkers encoding in the global multiclass maps (in the folder \texttt{All/}) while the "Label" column defines it for the individual multiclass maps (anatomical, vessels, bright, and red lesions).}
    \label{tab:semanticSegmentationLabels}
\end{table}

\paragraph{diagnosis.xls} is a spreadsheet (in Excel format) containing the DR and ME grades using the Canadian Teleophtalmology Guidelines system. The first worksheet \texttt{"Summary"} has 3 columns: \texttt{name} indexes each row with the image filename (without extensions), \texttt{ DR} stores the final grades of diabetic retinopathy, and \texttt{ ME} stores the final grades of macular edema. The following worksheets, \texttt{"DR"} and \texttt{"ME"}, present the same data under the column \texttt{name} and \texttt{consensus} and additionally provides the individual grades annotated by each retinologist under the columns \texttt{RetinologistA}, \texttt{RetinologistB} and \texttt{RetinologistC}. The last two worksheets, \texttt{Comment} and \texttt{UntranslatedComment}, are again indexed by \texttt{name} and contain the comments left by retinologists during labeling in their translated and original French versions. 

\paragraph{biomarkers-supplementaryInformation.zip} provides complementary information and data collected during the annotation process and published for transparency purposes. The Excel file \texttt{biomarkers\_annotation\_infos.xls} consists of four worksheets, one for every category of biomarkers: \texttt{bright}, \texttt{red}, \texttt{vessels}, and \texttt{anatomical}. For each category and each MAPLES-DR image identified by \texttt{name}, those tables show the \texttt{Retinologist} who performed the annotation, the translation of any \texttt{Comment} left by the retinologist (as well as the \texttt{Untranslated Comment} in French), the \texttt{Time} in second spent annotating and the annotation rank: \texttt{Annotation \#} (where 1 indicates the first image that was annotated and 200 indicates the last). The folder \texttt{preannotations/} and \texttt{annotations/} contains the original preannotated and annotated segmentation maps before rescaling and splitting into a train and test set. The folder \texttt{duplicates/} contains the segmentation maps and the diagnosis of the two duplicate images. Finally, the file \texttt{MESSIDOR-rois.yaml} provides the bounding rectangles in the MESSIDOR coordinate system (encoded as [top line, left column, height, width]) of the segmentation maps in the \texttt{preannotations/} and \texttt{annotations/} folders.

\section*{Technical Validation}

\subsection*{DR and ME grades: overview}
\begin{figure}[ht]
\centering
\includegraphics[width=0.65\linewidth]{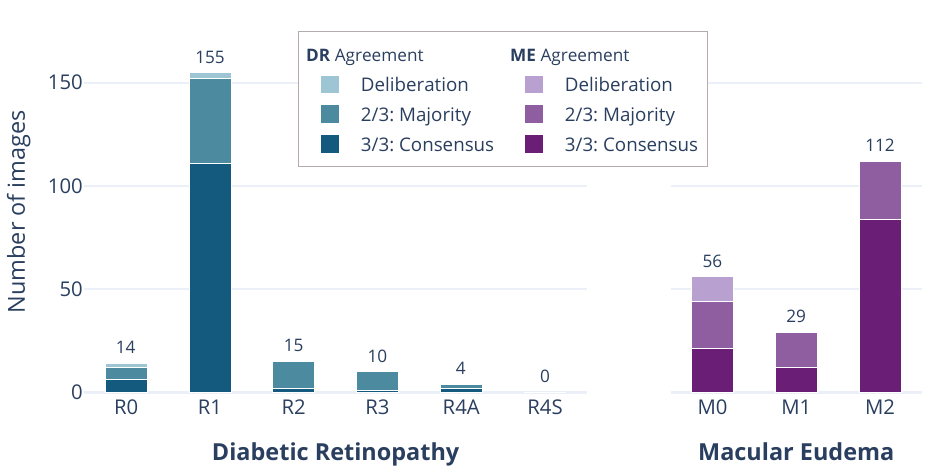}
\caption{Image counts for each DR and ME grade in MAPLES-DR and number of retinologists agreeing with the final grade. (Deliberation was required when all three retinologists were in disagreement).}
\label{fig:MAPLES-DR-dist}
\end{figure}

Of the 198 images selected from MESSIDOR, all were judged to be of sufficient quality to receive a DR grade. Only one was deemed insufficient for ME grading. Our team of retinologists diagnosed a large majority of images as stage \emph{R1: mild DR} and none as stage \emph{R4S: Stable Treated Proliferative} (cf. Figure~\ref{fig:MAPLES-DR-dist}). From this perspective, the discrepancy between the grade distribution published in MESSIDOR that we used to select the 200 images (cf. Figure\ref{fig:DR-dist}) and the MAPLES-DR grades may seem significant, but two factors must be considered to understand the validity of each dataset. First, the visualization and preprocessing tools available in our annotation platform promote high sensitivity of lesions detection. Assisted by these tools, MAPLES-DR retinologists identified enough microaneurysms in half of MESSIDOR R0 (healthy) images to reclassify them as R1. Second, to be diagnosed as R2 in MAPLES-DR, an image must contain at least 4 hemorrhages (according to international grading guidelines), but it only needs to contain 1 hemorrhage or 5 microaneurysms to receive the same grade in MESSIDOR. Due to this difference between the two grading systems, many images classified as R2 or R3 in MESSIDOR are reclassified as R1 in MAPLES-DR. The opposite trend is seen for the macular edema scores: Although most of the selected images were classified as M0 in MESSIDOR, many were reclassified as M2 in MAPLES-DR because the former grading focuses on exudates, while the latter grading also considers microaneurysms and hemorrhages within the macula. These results underline the impact of the grading systems on global pathological labels and validate \textit{a posteriori} our choice to reannotate the images in a system comparable to the international standard.

The DR and ME grading procedure requested a deliberation phase when all three retinologists were in disagreement. This step was only necessary for five DR grades and 12 ME grades. The deliberation always resulted in a low severity grade: R0, R1 or M0 (see Figure~\ref{fig:MAPLES-DR-dist}). By contrast, the majority of images were identically graded by the three retinologists (122/198 images for DR and 116/198 images for ME).

\subsection*{Retinal structures: overview}
The pixel-wise annotation of DR biomarkers was performed primarily by the three retinologists who also graded DR and ME. Each of them specialized in one category of biomarkers: vessels, red lesions, and bright lesions. The other four retinologists contributed mainly to the annotation of the optic disc and macula, as well as red and bright lesions for a small number of images. In total, the team dedicated 69 hours to review preannotated structures, erasing false positives, manually adding annotations that were missing from the preannotated maps, and correcting lesions that were correctly segmented but misclassified (e.g., microaneurysms instead of hemorrhages). 

\begin{figure}[ht]
\centering
\begin{subfigure}{.7425\textwidth}
        \includegraphics[width=\linewidth]{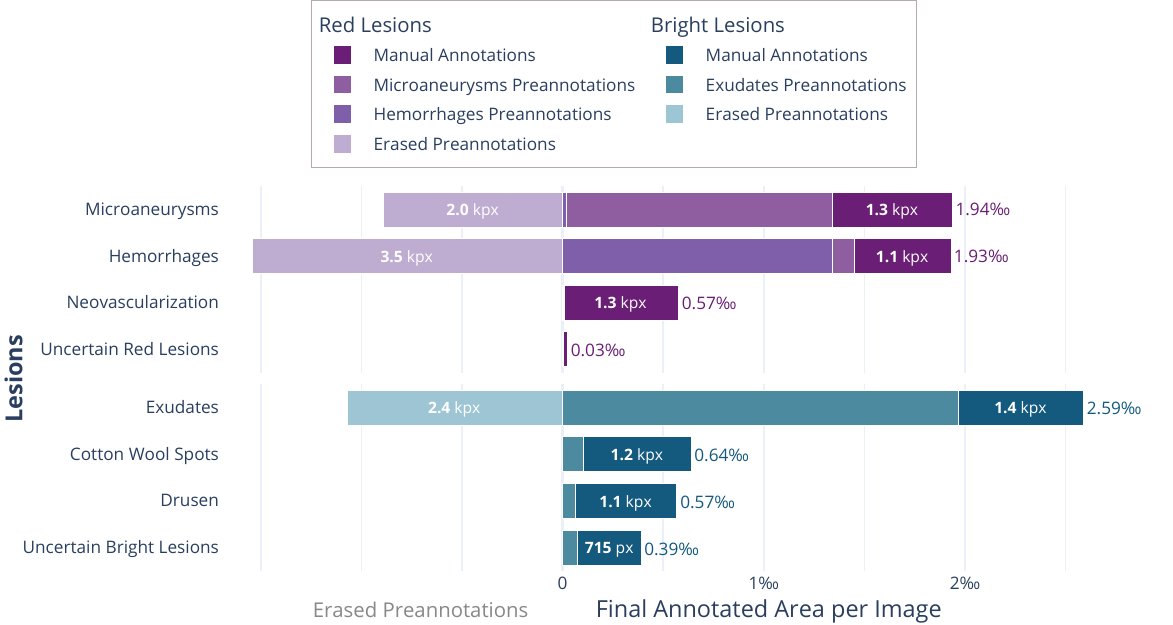}
    \end{subfigure}
    \begin{subfigure}{.2475\textwidth}
        \includegraphics[width=\linewidth]{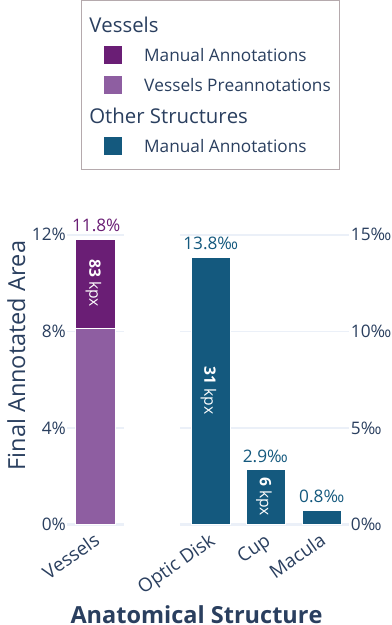}
    \end{subfigure}

\caption{Overview of the annotation work performed by the retinologists for each category of biomarkers: red lesions, bright lesions, vessels, and other anatomical structures. Manually added annotations appear darker while manually erased pre-\allowbreak annotations appear dimmer. Both are labeled with the average number of corrected pixels per image. The final area covered by each biomarker after correction is displayed as a fraction of the total image area of the dataset.}

\label{fig:diff}
\end{figure}

The macula and optic disc (including the cup) were the fastest to be annotated, with an average of 2 minutes per image, followed by vessel segmentation, with an average of 6 minutes per image. For the latter, preannotation appears to have been a valuable aid, since $69\%$ of the pixels found in the vascular maps published in MAPLES-DR were preannotated and less than $1.3\%$ of pixels from the initial preannotation maps were manually erased as false positive. However, the vessel labeling process still required substantial manual corrections: On average, an area of 83,000 pixels was manually added to each segmentation map (see Figure~\ref{fig:diff}).  In terms of image area, anatomical structures account for most of the MAPLES-DR annotations because, in addition to being typically larger than lesions, they are present in all images. The optic cup is the only exception: In five images, its boundaries were too uncertain for segmentation.

Despite covering only a small fraction of the MAPLES-DR images ($0.45\%$ of pixels for red lesions and $0.42\%$ for bright lesions), pathological structures required the greatest annotation effort. The time spent on each image varied greatly: from a few dozen seconds for healthy images to an hour for the most severe cases. Similarly to vessels, annotations of preannotated lesions (microaneurysms, hemorrages, and exudates) relied heavily on automatic segmentation. Only a quarter of the pixels were manually annotated, while the rest came directly from preannotated maps. However, unlike vessels, preannotation lesion maps contained more false positives and required a more careful review by retinologists to eliminate them. More than $50\%$ of the preannotated hemorrhages were erased, as were $40\%$ of the microaneurysms and $35\%$ of the exudates (see Figure~\ref{fig:diff}). In general, the annotation of pathological structures took an average of 6 minutes for bright lesions and 10 minutes for red lesions.

\begin{figure}[ht!]
\centering
\includegraphics[width=\linewidth]{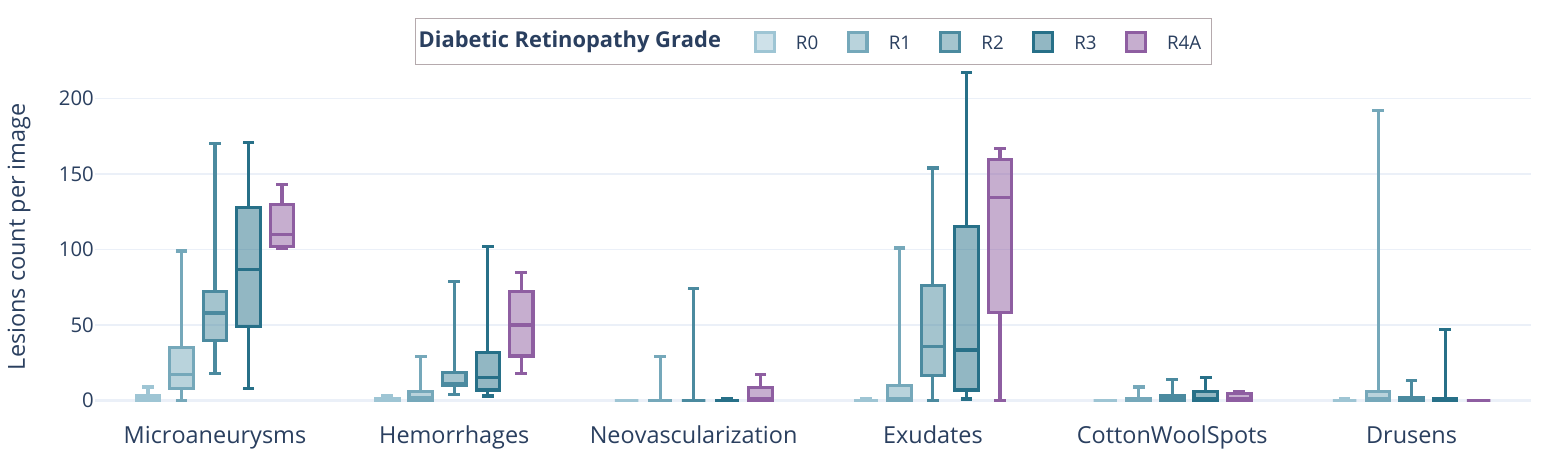}
\includegraphics[width=\linewidth]{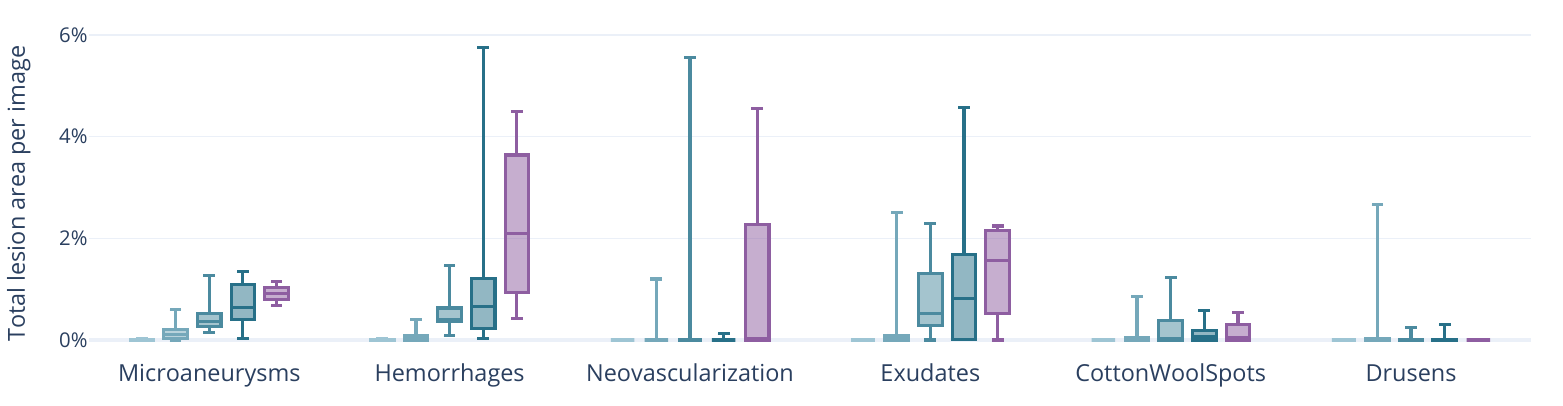}

\caption{Distribution of the number of individual lesions and their total area per image for each type of pathological structure, grouped by DR severity.}
\label{fig:pathological_dist}
\end{figure}

To assess the relevance of pathological structure annotations, we calculated the number of individual lesions per image, as well as their total area, and compared the distribution of both with respect to the severity of diabetic retinopathy (cf. Figure~\ref{fig:pathological_dist}). The general trend of these distributions matches the clinicial intuition: the number of lesions per image and the area they cover increase with the severity of the pathology. Furthermore, the distribution of individual lesions is in line with the clinical definition of each grade. The number of microaneurysms, characteristic of low-severity grades, increases substantially between R0 and R1, and again between R1 and R2. For hemorrhages and exudates, whose count distinguishes the more severe stages, the most significant transitions are between stages R1 and R2, and between stages R3 and R4A. Similarly, the average surface area covered by the neovessels increases between stages R3 and R4A. Such transitions are absent in the cotton wool spots and drusen distributions, as there are too few of them in MAPLES-DR to observe any trends ($<0.01\%$ of total pixels).

\subsection*{Retinal structures: semantic segmentation baseline}

The learnability of MAPLES-DR labels of retinal structures was tested on a semantic segmentation task by training a simple UNet model to jointly segment them all as a multiclass map. We used a straightforward training protocol that does not involve cross-validation, model ensembles, sophisticated regularizations, or extensive hyperparameter tuning. All these techniques would very likely significantly improve the performances obtained, but are outside the scope of this study. The model was trained with a learning rate of 0.003, using the Dice coefficient as the loss function and stochastic gradient descent (SGD) as the optimizer. To accelerate the training, the encoder's initial weights were pretrained on ImageNet. 

The semantic segmentation performances of the model trained on the MAPLES-DR training set and tested on its test set are summarized in Figure~\ref{fig:autoseg}. The model was able to successfully segment the four anatomical structures, as well as exudates, microaneurysms, and hemorrhages. Segmentation performances for these last two lesions may appear low but are, in fact, comparable to the scores obtained when learning to segment these lesions with other public databases. For example, the average segmentation precision of microaneurysms of a model trained and tested on MAPLES-DR (where microaneurysms are particularly well represented) is significantly higher than that obtained with a model trained and tested on IDRiD (see the competition leaderboard presented in Table~\ref{tab:pretraining}). Unfortunately, not all lesions provided in MAPLES-DR are as easily learnable. Our simple model was unable to provide satisfactory segmentation for CWS, drusen, and neovessels. Indeed, because MAPLES-DR contains few images of severe or proliferative DR, examples of these lesions are rare in its training set and even rarer in its test set.

\def\colSize{.22\textwidth}
\begin{figure}[ht]
    \centering
    \begin{subfigure}{\colSize}
        \includegraphics[width=\textwidth]{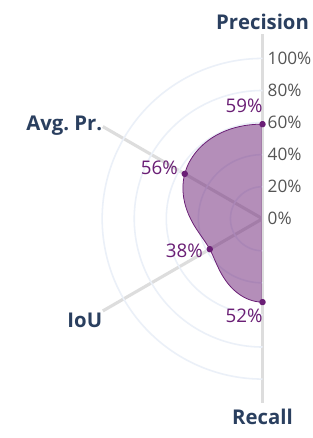}
        \subcaption{Microaneurysms}
    \end{subfigure}
    \begin{subfigure}{\colSize}
        \includegraphics[width=\textwidth]{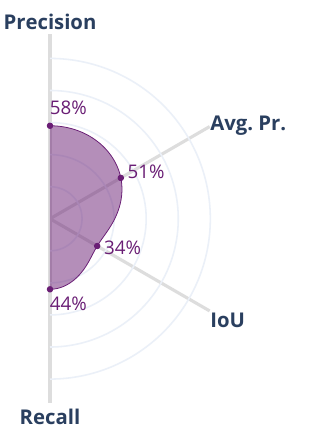}
        \subcaption{Hemorrhages}
    \end{subfigure}
    \begin{subfigure}{\colSize}
        \includegraphics[width=\textwidth]{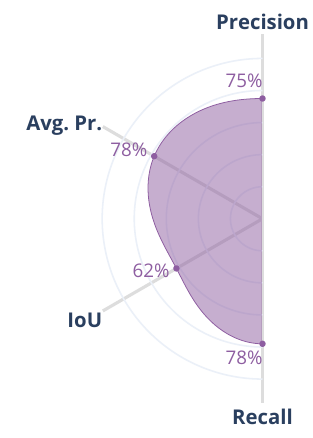}
        \subcaption{Exudates}
    \end{subfigure}
    
    \begin{subfigure}{\colSize}
        \includegraphics[width=\textwidth]{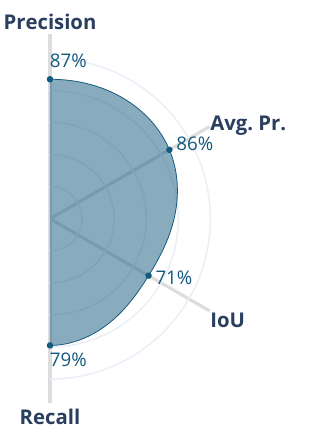}
        \subcaption{Vessels}
    \end{subfigure}
    \begin{subfigure}{\colSize}
        \includegraphics[width=\textwidth]{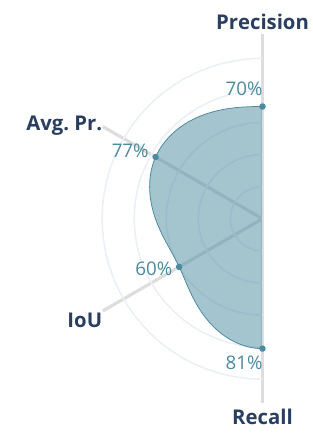}
        \subcaption{Optic Disc}
    \end{subfigure}
    \begin{subfigure}{\colSize}
        \includegraphics[width=\textwidth]{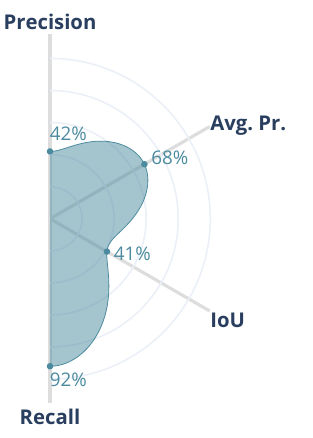}
        \subcaption{Optic Cup}
    \end{subfigure}
        \begin{subfigure}{\colSize}
        \includegraphics[width=\textwidth]{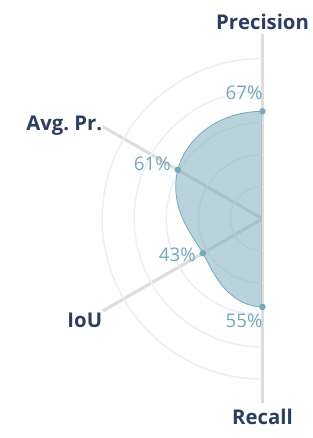}
        \subcaption{Macula}
    \end{subfigure}
    
    \caption{Summary of the semantic segmentation performances of a simple UNet model trained to segment all retinal structures available in MAPLES-DR.}
    \label{fig:autoseg}
\end{figure}

\subsection*{Retinal lesions: pretraining with MAPLES-DR}

As a final validation step, we illustrate how MAPLES-DR can be used in conjunction with other existing public datasets to improve segmentation performance. Various factors such as annotation styles, colorimetry, and image resolutions, can differ noticeably between datasets. Because of such distribution misalignments, it is often not relevant to test a model on one dataset while having it trained on another, or even on a combination of both. Transfer learning is generally a better choice to take advantage of features learned on one dataset to improve training on another. We tested the benefit of MAPLES-DR in this context by training two simple UNet models on the IDRiD retinal lesions dataset. While both models were initialized with weights pre-trained on ImageNet, one was also pretrained on the MAPLES-DR training set before being fine-tuned on IDRiD.  In both cases, we used the same basic training protocol as in the previous section. The resulting average precision scores for the segmentation of microaneurysms, hemorrhages, exudates, and cotton wool spots are presented in Table~\ref{tab:pretraining}, along with the IDRiD competition leaderboard.

\begin{table}[t]
    \centering
    \begin{tabular}{lccccc}
    \toprule
         Model & MA & HEM  & EX & CWS & Average \\
         \midrule
         \multicolumn{6}{c}{Competition Leaderboard} \\
         \midrule
         iFLYTEK-MIG  & \textbf{0.502}& 0.559 & 0.874 & 0.659 & \textbf{0.649}\\
         PATech & 0.474 & 0.649 & \textbf{0.885}& - & -\\
 VRT & 0.495 & 0.680 & 0.713 & \textbf{0.699}&0.647 \\
         \midrule
        UNet trained on IDRiD & 0.433  & 0.616 & 0.818 & 0.596 & 0.616 \\
        UNet pretrained on MAPLES-DR, finetuned on IDRiD & 0.432 & \textbf{0.685}& 0.816 & 0.661 & \textbf{0.649}\\
        \bottomrule
    \end{tabular}
    \caption{Effect of MAPLES-DR pretraining on the average precision of lesion segmentation on the IDRiD test set. }
    \label{tab:pretraining}
\end{table}

While pretraining with MAPLES-DR did not have much effect on microaneurysms and exudates, it significantly improved the segmentation of hemorrhages and cotton wool spots. Even though it was trained using a rudimentary method, the model reached an average score of 0.649 for all lesions, matching the best submission to the IDRiD challenge at the time it was held.

\section*{Usage Notes}
The fundus images corresponding to the diagnostic labels and biomarker segmentation maps of MAPLES-DR are the property of the MESSIDOR consortium and are freely available from \href{https://www.adcis.net/fr/logiciels-tiers/messidor-fr/}{the Consortium's website} . 

Regarding the lesion segmentation maps, we discourage the use of our drusen, CWS, and neovessel maps to evaluate models trained specifically to segment or detect those biomarkers. MAPLES-DR contains only a few examples of those structures (even fewer in its test set), and fundus imaging is not the clinical gold standard for detecting them. More generally, for the detection of neo-vessels in MAPLES-DR images, we recommend relying instead on the R4A grade or on the clinicians' comments when annotating vessels. 

We also wish to draw the reader's attention to the non-negligible number of bright lesions annotated as  \texttt{BrightUncertains} when they were clearly pathological but difficult to classify as drusen, exhudates or CWS. We recommend including them in the training and testing sets of models that segment or detect bright lesions without classifying them. For image-specific information on bright lesions annotated as uncertain, please consult the comments left by the annotators.

Finally, the MAPLES-DR segmentation maps were padded and resized to heterogeneous sizes and resolutions to match the original MESSIDOR fundus images. However, libraries specialized in training segmentation models commonly expect a standardized image size. We advise researchers who wish to use MAPLES-DR in such a context to crop and resize MESSIDOR fundus images with the same ROI we used to annotate them (the code described in the Code availability section provides Python examples).

\section*{Code availability}
The annotation platform code and documentation are freely available to any research team that needs it on the Github repository: \href{https://github.com/LIV4D/AnnotationPlatform}{LIV4D/AnnotationPlatform} .

The algorithms used to generate the preannotation maps cannot be made public as they were implemented with the Theano library, which is no longer maintained. For transparency purposes, the original preannotations are available in the supplementary information archive of MAPLES-DR. However, we have released up-to-date Python segmentation models (trained with MAPLES-DR and other public datasets) on two Github repositories: \href{https://github.com/gabriel-lepetitaimon/fundus-vessels-toolkit}{gabriel-lepetitaimon/fundus-vessels-toolkit} provides automatic segmentation and graph extraction of the retinal vasculature; \href{https://github.com/ClementPla/fundus-lesions-toolkit}{ClementPla/fundus-lesions-toolkit} provides automatic segmentation and visualization of retinal lesions. These models perform noticeably better than the ones originally used to preannotate MAPLES-DR.

Finally, the code to format the database and generate the figures in this article is available on the \href{https://github.com/LIV4D/MAPLES-DR}{LIV4D/MAPLES-DR} Github repository. A Python library that provides tools to simplify the use of MAPLES-DR in a machine learning context is also included.

\bibliographystyle{unsrtnat}
\bibliography{bibliography}

\begin{thebibliography}{38}
\providecommand{\natexlab}[1]{#1}
\providecommand{\url}[1]{\texttt{#1}}
\expandafter\ifx\csname urlstyle\endcsname\relax
  \providecommand{\doi}[1]{doi: #1}\else
  \providecommand{\doi}{doi: \begingroup \urlstyle{rm}\Url}\fi

\bibitem[Lanzetta et~al.(2020)Lanzetta, , Sarao, Scanlon, Barratt, Porta, Bandello, and Loewenstein]{Lanzetta2020}
Paolo Lanzetta, , Valentina Sarao, Peter~H. Scanlon, Jane Barratt, Massimo Porta, Francesco Bandello, and Anat Loewenstein.
\newblock Fundamental principles of an effective diabetic retinopathy screening program.
\newblock \emph{Acta Diabetologica}, 57\penalty0 (7):\penalty0 785--798, mar 2020.
\newblock \doi{10.1007/s00592-020-01506-8}.

\bibitem[Cheng(2013)]{Committee2013}
Alice Y~Y Cheng.
\newblock {Canadian Diabetes Association} 2013 clinical practice guidelines for the prevention and management of diabetes in {Canada}.
\newblock \emph{Canadian Journal of Diabetes}, 37\penalty0 (Suppl 1):\penalty0 S1--S3, 2013.

\bibitem[Hooper et~al.(2017)Hooper, Boucher, Cruess, Dawson, Delpero, Greve, Kozousek, Lam, and Maberley]{Hooper2017}
Philip Hooper, Marie~Carole Boucher, Alan Cruess, Keith~G. Dawson, Walter Delpero, Mark Greve, Vladimir Kozousek, Wai-Ching Lam, and David~A.L. Maberley.
\newblock Excerpt from the {Canadian Ophthalmological Society} evidence-based clinical practice guidelines for the management of diabetic retinopathy.
\newblock \emph{Canadian Journal of Ophthalmology}, 52:\penalty0 S45--S74, nov 2017.
\newblock \doi{10.1016/j.jcjo.2017.09.027}.

\bibitem[Egunsola et~al.(2021)Egunsola, Dowsett, Diaz, Brent, Rac, and Clement]{Egunsola2021}
Oluwaseun Egunsola, Laura~E. Dowsett, Ruth Diaz, Michael~H. Brent, Valeria Rac, and Fiona~M. Clement.
\newblock Diabetic retinopathy screening: A systematic review of qualitative literature.
\newblock \emph{Canadian Journal of Diabetes}, 45\penalty0 (8):\penalty0 725--733.e12, dec 2021.
\newblock \doi{10.1016/j.jcjd.2021.01.014}.

\bibitem[Avidor et~al.(2020)Avidor, Loewenstein, Waisbourd, and Nutman]{Avidor2020}
Daniel Avidor, Anat Loewenstein, Michael Waisbourd, and Amir Nutman.
\newblock Cost-effectiveness of diabetic retinopathy screening programs using telemedicine: a systematic review.
\newblock \emph{Cost Effectiveness and Resource Allocation}, 18\penalty0 (1), apr 2020.
\newblock \doi{10.1186/s12962-020-00211-1}.

\bibitem[Raman et~al.(2019)Raman, Srinivasan, Virmani, Sivaprasad, Rao, and Rajalakshmi]{raman2019fundus}
Rajiv Raman, Sangeetha Srinivasan, Sunny Virmani, Sobha Sivaprasad, Chetan Rao, and Ramachandran Rajalakshmi.
\newblock Fundus photograph-based deep learning algorithms in detecting diabetic retinopathy.
\newblock \emph{Eye}, 33\penalty0 (1):\penalty0 97--109, 2019.

\bibitem[Dugas et~al.(2015)Dugas, Jorge, and Cukierski]{diabetic-retinopathy-detection}
Emma Dugas, Jared Jorge, and Will Cukierski.
\newblock Diabetic retinopathy detection.
\newblock \url{https://kaggle.com/competitions/diabetic-retinopathy-detection}, 2015.

\bibitem[Decenci{\`e}re et~al.(2014)Decenci{\`e}re, Zhang, Cazuguel, Lay, Cochener, Trone, Gain, Ordonez, Massin, Erginay, et~al.]{MESSIDOR}
Etienne Decenci{\`e}re, Xiwei Zhang, Guy Cazuguel, Bruno Lay, B{\'e}atrice Cochener, Caroline Trone, Philippe Gain, Richard Ordonez, Pascale Massin, Ali Erginay, et~al.
\newblock Feedback on a publicly distributed image database: the {MESSIDOR} database.
\newblock \emph{Image Analysis \& Stereology}, 33\penalty0 (3):\penalty0 231--234, 2014.

\bibitem[Gulshan et~al.(2019)Gulshan, Rajan, Widner, Wu, Wubbels, Rhodes, Whitehouse, Coram, Corrado, Ramasamy, Raman, Peng, and Webster]{Gulshan2019}
Varun Gulshan, Renu~P. Rajan, Kasumi Widner, Derek Wu, Peter Wubbels, Tyler Rhodes, Kira Whitehouse, Marc Coram, Greg Corrado, Kim Ramasamy, Rajiv Raman, Lily Peng, and Dale~R. Webster.
\newblock Performance of a deep-learning algorithm vs manual grading for detecting diabetic retinopathy in {India}.
\newblock \emph{{JAMA} Ophthalmology}, 137\penalty0 (9):\penalty0 987, sep 2019.
\newblock \doi{10.1001/jamaophthalmol.2019.2004}.

\bibitem[Phillips et~al.(2020)Phillips, Hahn, Fontana, Broniatowski, and Przybocki]{phillips2020four}
P~Jonathon Phillips, Carina~A Hahn, Peter~C Fontana, David~A Broniatowski, and Mark~A Przybocki.
\newblock Four principles of explainable artificial intelligence.
\newblock Technical Report NISTIR 8312, National Institute of Standards and Technology, Gaithersburg, Maryland, 2020.

\bibitem[Wilkinson et~al.(2003)Wilkinson, Ferris~III, Klein, Lee, Agardh, Davis, Dills, Kampik, Pararajasegaram, Verdaguer, et~al.]{wilkinson2003proposed}
Charles~P Wilkinson, Frederick~L Ferris~III, Ronald~E Klein, Paul~P Lee, Carl~David Agardh, Matthew Davis, Diana Dills, Anselm Kampik, R~Pararajasegaram, Juan~T Verdaguer, et~al.
\newblock Proposed international clinical diabetic retinopathy and diabetic macular edema disease severity scales.
\newblock \emph{Ophthalmology}, 110\penalty0 (9):\penalty0 1677--1682, 2003.

\bibitem[Zachariah et~al.(2015)Zachariah, Wykes, and Yorston]{Zachariah2015}
Sonia Zachariah, William Wykes, and David Yorston.
\newblock Grading diabetic retinopathy ({DR}) using the {Scottish} grading protocol.
\newblock \emph{Community Eye Health}, 28:\penalty0 72--73, 2015.
\newblock ISSN 0953-6833.

\bibitem[Boucher et~al.(2020)Boucher, Qian, Brent, Wong, Sheidow, Duval, Kherani, Dookeran, Maberley, Samad, and Chaudhary]{Boucher2020}
M.C. Boucher, J.~Qian, M.H. Brent, D.T. Wong, T.~Sheidow, R.~Duval, A.~Kherani, R.~Dookeran, D.~Maberley, A.~Samad, and V.~Chaudhary.
\newblock Evidence-based {Canadian} guidelines for tele-retina screening for diabetic retinopathy: recommendations from the {Canadian Retina Research Network (CR2N) Tele-Retina Steering Committee}.
\newblock \emph{Canadian Journal of Ophthalmology}, 55\penalty0 (1):\penalty0 14--24, feb 2020.
\newblock \doi{10.1016/j.jcjo.2020.01.001}.

\bibitem[Staal et~al.(2004)Staal, Abr{\`a}moff, Niemeijer, Viergever, and Van~Ginneken]{DRIVE}
Joes Staal, Michael~D Abr{\`a}moff, Meindert Niemeijer, Max~A Viergever, and Bram Van~Ginneken.
\newblock Ridge-based vessel segmentation in color images of the retina.
\newblock \emph{IEEE Transactions on Medical Imaging}, 23\penalty0 (4):\penalty0 501--509, 2004.

\bibitem[Budai et~al.(2013)Budai, Bock, Maier, Hornegger, Michelson, et~al.]{HRF}
Attila Budai, R{\"u}diger Bock, Andreas Maier, Joachim Hornegger, Georg Michelson, et~al.
\newblock Robust vessel segmentation in fundus images.
\newblock \emph{International Journal of Biomedical Imaging}, 2013, 2013.

\bibitem[Jin et~al.(2022)Jin, Huang, Zhou, Li, Yan, Sun, Zhang, Wang, and Ye]{FIVES}
Kai Jin, Xingru Huang, Jingxing Zhou, Yunxiang Li, Yan Yan, Yibao Sun, Qianni Zhang, Yaqi Wang, and Juan Ye.
\newblock Fives: A fundus image dataset for artificial intelligence based vessel segmentation.
\newblock \emph{Scientific Data}, 9\penalty0 (1):\penalty0 475, 2022.

\bibitem[Lyu et~al.(2022)Lyu, Cheng, and Zhang]{RETA}
Xingzheng Lyu, Li~Cheng, and Sanyuan Zhang.
\newblock The reta benchmark for retinal vascular tree analysis.
\newblock \emph{Scientific Data}, 9\penalty0 (1):\penalty0 397, 2022.

\bibitem[Orlando et~al.(2020)Orlando, Fu, Breda, Van~Keer, Bathula, Diaz-Pinto, Fang, Heng, Kim, Lee, et~al.]{REFUGE}
Jos{\'e}~Ignacio Orlando, Huazhu Fu, Jo{\~a}o~Barbosa Breda, Karel Van~Keer, Deepti~R Bathula, Andr{\'e}s Diaz-Pinto, Ruogu Fang, Pheng-Ann Heng, Jeyoung Kim, JoonHo Lee, et~al.
\newblock Refuge challenge: A unified framework for evaluating automated methods for glaucoma assessment from fundus photographs.
\newblock \emph{Medical Image Analysis}, 59:\penalty0 101570, 2020.

\bibitem[Kovalyk et~al.(2022)Kovalyk, Morales-S{\'a}nchez, Verd{\'u}-Monedero, Sell{\'e}s-Navarro, Palaz{\'o}n-Cabanes, and Sancho-G{\'o}mez]{PAPILA}
Oleksandr Kovalyk, Juan Morales-S{\'a}nchez, Rafael Verd{\'u}-Monedero, Inmaculada Sell{\'e}s-Navarro, Ana Palaz{\'o}n-Cabanes, and Jos{\'e}-Luis Sancho-G{\'o}mez.
\newblock Papila: Dataset with fundus images and clinical data of both eyes of the same patient for glaucoma assessment.
\newblock \emph{Scientific Data}, 9\penalty0 (1):\penalty0 291, 2022.

\bibitem[Kumar et~al.(2023)Kumar, Seelamantula, Gagan, Kamath, Kuzhuppilly, Vivekanand, Gupta, and Patil]{CHAKSU}
JR~Harish Kumar, Chandra~Sekhar Seelamantula, JH~Gagan, Yogish~S Kamath, Neetha~IR Kuzhuppilly, U~Vivekanand, Preeti Gupta, and Shilpa Patil.
\newblock Ch{\'a}kṣu: A glaucoma specific fundus image database.
\newblock \emph{Scientific Data}, 10\penalty0 (1):\penalty0 70, 2023.

\bibitem[Lin et~al.(2020)Lin, Li, Huang, Cheng, Xia, Wang, Yuan, and Tang]{SUSTECH}
Li~Lin, Meng Li, Yijin Huang, Pujin Cheng, Honghui Xia, Kai Wang, Jin Yuan, and Xiaoying Tang.
\newblock The {SUSTech}-{SYSU} dataset for automated exudate detection and diabetic retinopathy grading.
\newblock \emph{Scientific Data}, 7\penalty0 (1), nov 2020.
\newblock \doi{10.1038/s41597-020-00755-0}.

\bibitem[Decenci\`ere et~al.(2013)Decenci\`ere, Cazuguel, Zhang, Thibault, Klein, Meyer, Marcotegui, Quellec, Lamard, Danno, Elie, Massin, Viktor, Erginay, Laÿ, and Chabouis]{Decenciere2013}
E.~Decenci\`ere, G.~Cazuguel, X.~Zhang, G.~Thibault, J.-C. Klein, F.~Meyer, B.~Marcotegui, G.~Quellec, M.~Lamard, R.~Danno, D.~Elie, P.~Massin, Z.~Viktor, A.~Erginay, B.~Laÿ, and A.~Chabouis.
\newblock Teleophta: Machine learning and image processing methods for teleophthalmology.
\newblock \emph{IRBM}, 34\penalty0 (2):\penalty0 196 -- 203, 2013.
\newblock ISSN 1959-0318.
\newblock \doi{10.1016/j.irbm.2013.01.010}.
\newblock URL \url{http://www.sciencedirect.com/science/article/pii/S1959031813000237}.
\newblock Special issue : ANR TECSAN : Technologies for Health and Autonomy.

\bibitem[Zhou et~al.(2020)Zhou, Wang, Huang, Cui, and Shao]{FGADR}
Yi~Zhou, Boyang Wang, Lei Huang, Shanshan Cui, and Ling Shao.
\newblock A benchmark for studying diabetic retinopathy: segmentation, grading, and transferability.
\newblock \emph{IEEE Transactions on Medical Imaging}, 40\penalty0 (3):\penalty0 818--828, 2020.

\bibitem[Wei et~al.(2021)Wei, Li, Yu, Zhang, Zhang, Hu, Mo, Gong, Chen, Ding, et~al.]{RetinalLesions}
Qijie Wei, Xirong Li, Weihong Yu, Xiao Zhang, Yongpeng Zhang, Bojie Hu, Bin Mo, Di~Gong, Ning Chen, Dayong Ding, et~al.
\newblock Learn to segment retinal lesions and beyond.
\newblock In \emph{2020 25th International Conference on Pattern Recognition (ICPR)}, pages 7403--7410. IEEE, 2021.

\bibitem[Porwal et~al.(2018)Porwal, Pachade, Kamble, Kokare, Deshmukh, Sahasrabuddhe, and Meriaudeau]{IDRID}
Prasanna Porwal, Samiksha Pachade, Ravi Kamble, Manesh Kokare, Girish Deshmukh, Vivek Sahasrabuddhe, and Fabrice Meriaudeau.
\newblock Indian diabetic retinopathy image dataset ({IDRiD}): a database for diabetic retinopathy screening research.
\newblock \emph{Data}, 3\penalty0 (3):\penalty0 25, 2018.

\bibitem[Li et~al.(2019)Li, Gao, Wang, Guo, Liu, and Kang]{DDR}
Tao Li, Yingqi Gao, Kai Wang, Song Guo, Hanruo Liu, and Hong Kang.
\newblock Diagnostic assessment of deep learning algorithms for diabetic retinopathy screening.
\newblock \emph{Information Sciences}, 501:\penalty0 511--522, 2019.
\newblock ISSN 0020-0255.
\newblock \doi{10.1016/j.ins.2019.06.011}.
\newblock URL \url{https://www.sciencedirect.com/science/article/pii/S0020025519305377}.

\bibitem[Cao et~al.(2022)Cao, Felfeli, Merritt, and Brent]{Cao2022}
Jessica Cao, Tina Felfeli, Rebecca Merritt, and Michael~H. Brent.
\newblock Sociodemographics associated with risk of diabetic retinopathy detected by tele-ophthalmology: 5-year results of the toronto tele-retinal screening program.
\newblock \emph{Canadian Journal of Diabetes}, 46\penalty0 (1):\penalty0 26--31, feb 2022.
\newblock \doi{10.1016/j.jcjd.2021.05.001}.

\bibitem[Sasongko et~al.(2011)Sasongko, Wong, Nguyen, Cheung, Shaw, and Wang]{sasongkoRetinalVascularTortuosity2011}
M.~B. Sasongko, T.~Y. Wong, T.~T. Nguyen, C.~Y. Cheung, J.~E. Shaw, and J.~J. Wang.
\newblock Retinal vascular tortuosity in persons with diabetes and diabetic retinopathy.
\newblock \emph{Diabetologia}, 54\penalty0 (9):\penalty0 2409--2416, September 2011.
\newblock ISSN 1432-0428.
\newblock \doi{10.1007/s00125-011-2200-y}.

\bibitem[Purves et~al.(2001)Purves, Augustine, Fitzpatrick, Katz, LaMantia, McNamara, and Williams]{purvesRetinotopicRepresentationVisual2001}
Dale Purves, George~J. Augustine, David Fitzpatrick, Lawrence~C. Katz, Anthony-Samuel LaMantia, James~O. McNamara, and S.~Mark Williams.
\newblock The {{Retinotopic Representation}} of the {{Visual Field}}.
\newblock In \emph{Neuroscience}. {Sinauer Associates}, 2nd edition, 2001.

\bibitem[Playout et~al.(2019)Playout, Duval, and Cheriet]{playout2019novel}
Cl{\'e}ment Playout, Renaud Duval, and Farida Cheriet.
\newblock A novel weakly supervised multitask architecture for retinal lesions segmentation on fundus images.
\newblock \emph{IEEE Transactions on Medical Imaging}, 38\penalty0 (10):\penalty0 2434--2444, 2019.

\bibitem[Lepetit-Aimon et~al.(2018)Lepetit-Aimon, Duval, and Cheriet]{Lepetit2018}
Gabriel Lepetit-Aimon, Renaud Duval, and Farida Cheriet.
\newblock Large receptive field fully convolutional network for semantic segmentation of retinal vasculature in fundus images.
\newblock In \emph{Computational Pathology and Ophthalmic Medical Image Analysis}, pages 201--209. Springer, 2018.

\bibitem[R{\"a}dsch et~al.(2023)R{\"a}dsch, Reinke, Weru, Tizabi, Schreck, Kavur, Pekdemir, Ro{\ss}, Kopp-Schneider, and Maier-Hein]{radsch2023labelling}
Tim R{\"a}dsch, Annika Reinke, Vivienn Weru, Minu~D Tizabi, Nicholas Schreck, A~Emre Kavur, B{\"u}nyamin Pekdemir, Tobias Ro{\ss}, Annette Kopp-Schneider, and Lena Maier-Hein.
\newblock Labelling instructions matter in biomedical image analysis.
\newblock \emph{Nature Machine Intelligence}, 5\penalty0 (3):\penalty0 273--283, 2023.

\bibitem[Sechidis et~al.(2011)Sechidis, Tsoumakas, and Vlahavas]{sechidisStratificationMultilabelData2011}
Konstantinos Sechidis, Grigorios Tsoumakas, and Ioannis Vlahavas.
\newblock On the stratification of multi-label data.
\newblock In \emph{Machine Learning and Knowledge Discovery in Databases}, pages 145--158, {Berlin, Heidelberg}, 2011. Springer.
\newblock ISBN 978-3-642-23808-6.
\newblock \doi{10.1007/978-3-642-23808-6_10}.

\bibitem[Ruamviboonsuk et~al.(2006)Ruamviboonsuk, Teerasuwanajak, Tiensuwan, and Yuttitham]{Ruamviboonsuk2006}
Paisan Ruamviboonsuk, Khemawan Teerasuwanajak, Montip Tiensuwan, and Kanokwan Yuttitham.
\newblock Interobserver agreement in the interpretation of single-field digital fundus images for diabetic retinopathy screening.
\newblock \emph{Ophthalmology}, 113\penalty0 (5):\penalty0 826--832, may 2006.
\newblock \doi{10.1016/j.ophtha.2005.11.021}.

\bibitem[Bragge et~al.(2011)Bragge, Gruen, Chau, Forbes, and Taylor]{bragge2011}
Peter Bragge, Russell~L Gruen, Marisa Chau, Andrew Forbes, and Hugh~R Taylor.
\newblock Screening for presence or absence of diabetic retinopathy: a meta-analysis.
\newblock \emph{Archives of Ophthalmology}, 129\penalty0 (4):\penalty0 435--444, 2011.

\bibitem[Shi et~al.(2015)Shi, Wu, Dong, Jiang, Lu, and Shi]{shi2015}
Lili Shi, Huiqun Wu, Jiancheng Dong, Kui Jiang, Xiting Lu, and Jian Shi.
\newblock Telemedicine for detecting diabetic retinopathy: a systematic review and meta-analysis.
\newblock \emph{British Journal of Ophthalmology}, 99\penalty0 (6):\penalty0 823--831, 2015.

\bibitem[Krause et~al.(2018)Krause, Gulshan, Rahimy, Karth, Widner, Corrado, Peng, and Webster]{Krause2018}
Jonathan Krause, Varun Gulshan, Ehsan Rahimy, Peter Karth, Kasumi Widner, Greg~S. Corrado, Lily Peng, and Dale~R. Webster.
\newblock Grader variability and the importance of reference standards for evaluating machine learning models for diabetic retinopathy.
\newblock \emph{Ophthalmology}, 125\penalty0 (8):\penalty0 1264--1272, aug 2018.
\newblock \doi{10.1016/j.ophtha.2018.01.034}.

\bibitem[Teoh et~al.(2023)Teoh, Wong, Xiao, Wong, Zhao, Chan, Yuen, Naing, Yogesan, and Koh]{teoh2023variability}
Chin~Sheng Teoh, Kah~Hie Wong, Di~Xiao, Hung~Chew Wong, Paul Zhao, Hwei~Wuen Chan, Yew~Sen Yuen, Thet Naing, Kanagasingam Yogesan, and Victor Teck~Chang Koh.
\newblock Variability in grading diabetic retinopathy using retinal photography and its comparison with an automated deep learning diabetic retinopathy screening software.
\newblock \emph{Healthcare}, 11\penalty0 (12):\penalty0 1697, 2023.
\newblock \doi{10.3390/healthcare11121697}.

\end{thebibliography}


\section*{Acknowledgements}

This study was funded by the Natural Science and Engineering Research Council of Canada as well as Diabetes Action Canada and FROUM (Fonds de recherche en ophtalmologie de l'Université de Montréal). We thank Philippe Debanné for editing this manuscript and  Emmanuelle Richer, Zacharie Legault, and Fantin Girard for their valuable input on the technical validation and figures. 

\section*{Author contributions statement}

M.C.B and F.C conceptualized and designed the study.
G.L.A and C.P. designed and supervised the development of the annotation plateform.
M.C.B and R.D. and M.B. annotated the dataset.
G.L.A supervised the annotation campaigns, documented its procedure, and finally exported and analyzed the dataset.
C.P. performed and analyzed the trainability experiments.
G.L.A drafted the initial manuscript.
All authors reviewed and approved the final manuscript as submitted.

\section*{Competing interests} 
The authors have no competing interest to declare

\end{document}